\documentclass[preprint,12pt]{elsarticle}

\usepackage{hyperref}

\usepackage{amssymb,amsmath}
\graphicspath{{Figs/}}

\newcommand{\F}{F_\pi}
\allowdisplaybreaks

\begin{document}
\begin{frontmatter}
\title{Charge asymmetry in $e^+e^-\to \pi^+\pi^-$ process.}
\author{Fedor Ignatov}
\author{Roman N. Lee}

%\affiliation{organization={Budker Institute of Nuclear Physics, SB RAS},
%         addressline={},
%         city={Novosibirsk},
%         postcode={630090},
%         state={},
%         country={Russia}} 
% \affiliation{, Novosibirsk, 630090, Russia}
\address{Budker Institute of Nuclear Physics, SB RAS, Novosibirsk, 630090, Russia}
%\maketitle

\begin{abstract}
    We consider the charge asymmetry in the differential cross section of the process $e^+e^- \rightarrow \pi^+\pi^-$. Experimental data show large deviations from the results of the available theoretical calculations. Motivated by this fact, we revisit the contribution of the two photon exchange diagrams and find the origin of discrepancy in an oversimplified account of the pion internal structure in the theoretical calculations. We present a natural and simple approach that accounts for the pion structure in a more consistent way and find remarkable agreement with the experimental data.
\end{abstract}
\end{frontmatter}

\section{Introduction}

The $e^+e^-\to\pi^+\pi^-$ process is a dominant channel of the hadron production in the energy range below $\sqrt{s}<1$~GeV. The experimental total hadron production cross-section defines, via dispersion relation, the contribution of hadron vacuum polarization to different physical quantities, in particular, to the effective fine structure constant $\alpha_{eff}(E)$, Ref.  \cite{Jegerlehner:2006ju}, and to the anomalous magnetic moment of the muon $a_{\mu}=(g_{\mu}-2)/2$, Refs. \cite{Keshavarzi:2018mgv,Davier:2019can}. There is a long-standing $3\div4\sigma$ discrepancy between the experimental measurement and the Standard Model prediction for  $a_{\mu}$ which attracts a lot of attention in the context of searches of New Physics.
The $\pi^+\pi^-$ channel provides a major part of the hadronic contribution to $a_{\mu}$ being  $(506.0 \pm 3.4) \cdot 10^{-10}$. It also determines the overall theoretical precision $4.3\cdot 10^{-10}$ of the $g-2$ muon anomaly \cite{Aoyama:2020ynm}.
The final precision of the underway Fermilab experiment \cite{Muong-2:2015xgu} is expected
to be $\Delta a ^{exp}_{\mu}[E989] \approx 1.6 \cdot 10^{-10}$. In order to match this experimental precision from theoretical side, it is necessary to have 0.2\% overall systematic accuracy in the $\pi\pi$  channel.

In the present paper we consider the charge asymmetry in $e^+e^-\to \pi^+\pi^-$ process. In contrast to the total cross section, the charge asymmetry is appears entirely due to the radiative corrections. In the leading order this asymmetry is made up of two contributions. The virtual contribution comes from the interference of the Born diagram and the diagram with double photon exchange between lepton line and pion line, see Fig. \ref{fig:2ph_diagram}. The real radiation contribution to the asymmetry comes from the interference of initial state radiation (ISR) diagrams and  final state radiation (FSR) diagrams.  It depends on the experimental cuts which restrict the momentum of the emitted photon. Note that only the sum of these two contributions is finite while each of them is infrared divergent. Then, similar to Ref. \cite{Arbuzov:2020foj}, we write the charge-odd part of the differential cross section $\frac{d\sigma_{a}}{dc}$ in the form
\begin{equation}
    \frac{d\sigma_{a}}{dc}=\frac{d\sigma_{a}^V}{dc}+\frac{d\sigma_{a}^S(\omega_0)}{dc}+\frac{d\sigma_{a}^H(\omega_0)}{dc}\stackrel{\text{def}}{=}\frac{d\sigma_{0}}{dc}\left[\delta_{a}^V+\delta_{a}^S(\omega_0)\right]+\frac{d\sigma_{a}^H(\omega_0)}{dc}
\end{equation}
Here $c=\cos\theta$ with $\theta$ being the angle between the momenta of $\pi^-$ and electron,  $\frac{d\sigma_{a}^V}{dc}$ is the contribution to the cross section which comes from the interference between the Born diagram and the double photon exchange diagrams,  $\frac{d\sigma_{a}^S(\omega_0)}{dc}$ is the real radiation contribution which comes from the ISR-FSR interference with emitted photon having energy below $\omega_0$ in center-of-mass frame, and $\frac{d\sigma_{a}^H(\omega_0)}{dc}$ is the similar contribution from the photons with energy above $\omega_0$. It is this last contribution which depends on experimental cuts.

The Born cross section is
\begin{equation}
    \frac{d\sigma_{0}}{dc}=\frac{\pi\alpha^2}{4s}\beta^3(1-c^2)|\F(s)|^2,
\end{equation}
where $\beta=\sqrt{1-4m_\pi^2/s}$ is the velocity of $\pi$-meson.

\section{Conventional prescription for double photon exchange radiative corrections}
The crucial issue of any channel measurement in $e^+e^-$ annihilation is the account of radiative corrections. 
Of course, for hadronic channels, the calculation of the radiative corrections necessarily relies on some model. 
This is because hadrons are not point-like particles, and the internal structure of hadrons is governed by strong interactions which do not allow for perturbative treatment. 
For the process $e^+e^-\to\pi^+\pi^-(\gamma)$ the account of the pion internal structure traditionally has been reduced to the multiplication of the amplitude in scalar QED (sQED) with point-like pion by the pion form factor $\F(q^2=s)$, see, in particular, recent paper \cite{Arbuzov:2020foj}.
While this prescription is correct for the tree amplitude of $e^+e^-\to\pi^+\pi^-(\gamma)$ when the emitted photon is soft, it should be critically scrutinized for other kinematic regions as well as for other radiative corrections. In particular, there is no reason to expect that this receipt works well for the diagrams with double photon exchange between electron and pion lines. 

One possible argument in favor of this prescription is that it secures the cancellation of infrared divergences between the virtual correction and the real radiative correction. Technically, this cancellation is trivial as within this approach the inclusive cross section is obtained by multiplying the inclusive cross section in sQED, already IR finite, by $|\F(q^2)|^2$. However, it is instructive to explain, in a different way, why this simple prescription correctly accounts for the contribution of soft region responsible for IR divergence. 
\begin{figure}
    \centering
    \includegraphics[width=0.5\linewidth]{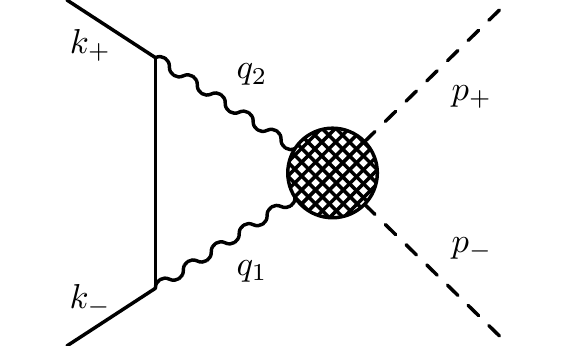}
    \caption{Double photon exchange contribution.}
    \label{fig:2ph_diagram}
\end{figure}

\begin{figure}
    \centering
    \includegraphics[width=\linewidth]{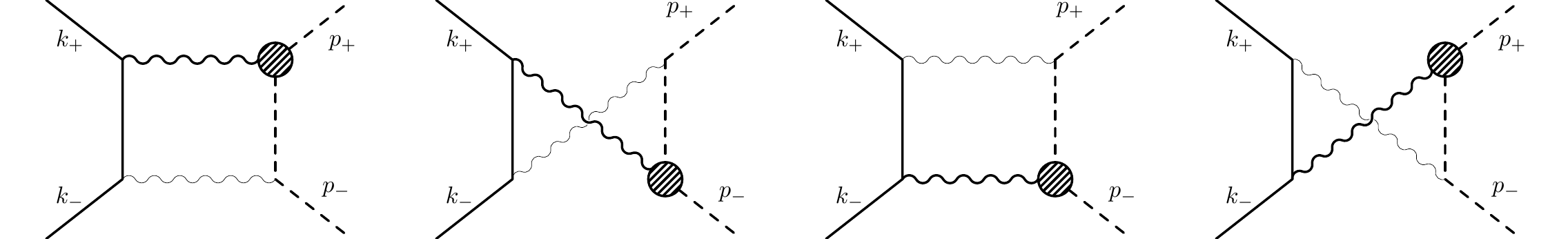}
    \caption{Schematic picture of soft photon contribution. Thin wavy line carries momentum whose components are all small.}
    \label{fig:soft_hard_diagrams}
\end{figure}

Let us consider the double photon exchange amplitude in Fig. \ref{fig:2ph_diagram}.
Infrared divergences come from two symmetric integration regions over loop momentum: $(q_1\to 0,\ q_2=q-q_1\to q)$ and $(q_2\to 0,\ q_1=q-q_2\to q)$. Let us consider, e.g., the first region. The soft photon with large wavelength does not ``see'' the internal structure of the pion. The IR divergent contributions can be depicted as in Fig. \ref{fig:soft_hard_diagrams}. Soft photon (thin wavy lines) is attached to external legs and the divergence comes from the propagators of slightly virtual pion and electron, corresponding to internal dashed and solid lines, respectively. As this pion is close to mass shell, the blob corresponds to the same factor which appears in the Born cross section --- to the form factor $\F(q_2^2)\approx \F(q^2)$.  
Thus, the traditional prescription --- multiplication of sQED amplitude  by $\F(q^2)$ --- correctly accounts the contribution of integration region responsible for IR divergence. However there is no reason to expect that hard region is also decently described within this approach. 

Let us discuss qualitatively what is going on when we gradually increase the soft photon momentum $q_1$. It it clear that this photon starts to ``see'' the internal structure of the pion. What is the characteristic scale of $q_1$ when the internal structure becomes essential? Naturally, this is the same scale at which the pion form factor essentially deviates from $1$, i.e., $|q_1|\sim m_{\rho}\sim 800$ MeV. What is going on when we have not yet reached this scale, $|q_1|\ll m_\rho$? We can still use the approximation in Fig. \ref{fig:soft_hard_diagrams}, however there is a subtlety here: the form factor in these diagrams depends on $q_2^2=(q-q_1)^2$ rather than on $q^2$. The difference between $\F(q_2^2)$ and $\F(q^2)$ might be essential when $q^2$ is in the vicinity of $m_\rho^2$. In this region the form factor changes rather rapidly with the characteristic scale $\delta q\sim \Gamma_{\rho}\sim 150$ MeV. Of course these qualitative considerations should be taken with a grain of salt if we recall that $\Gamma_{\rho}$ is only about $5$ times smaller than $m_\rho$. However one can expect that a model which somehow accounts these considerations will describe the asymmetry better than the traditional approach in the vicinity of $\sqrt{s}=m_{\rho}$. We shall see below that this is indeed the case. In the next section we consider a natural model with these properties.

\section{Generalized vector dominance in a loop}

A natural idea to improve the description of the two-photon contribution is to use the following modification of sQED Feynman rules:
\begin{align}
    \vcenter{\hbox{\includegraphics{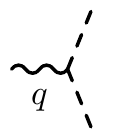}}}\longrightarrow &\quad
    \vcenter{\hbox{\includegraphics{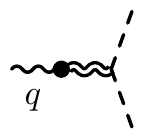}}}=\F(q^2)\times \vcenter{\hbox{\includegraphics{GVDff2}}}\nonumber\\
    \vcenter{\hbox{\includegraphics{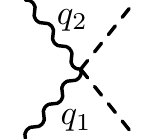}}}\longrightarrow &\quad
    \vcenter{\hbox{\includegraphics{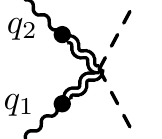}}}=\F(q_1^2)\F(q_2^2)\times\!\!\!\!\!\!\! \vcenter{\hbox{\includegraphics{GVDsg2}}}
    \label{eq:FFsQED}
\end{align}%%
As these graphical notations hint, this modification is inspired by the generalized vector dominance model  (GVD) \cite{Sakurai:1972wk}. In particular, we represent the pion form-factor as a sum of terms
\begin{equation}
    \F(q^2)=\sum_{v=1}^{n} a_v \frac{\Lambda_v}{\Lambda_v-q^2},\label{eq:FF}    
\end{equation}
where $\Lambda_v=m_v^2-im_v\Gamma_v$ is a complex parameter defined via mass and width of the vector meson $v$, and $a_v$ are some numbers. From the identity $\F(0)=1$ we have $\sum_{v} a_v=1$. 
Each term in the sum corresponds to the vector meson propagator multiplied by the coupling constant of $\gamma v$ vertex.
If we allow for sufficiently many terms in $\sum_v$ and treat the parameters $\Lambda_v$ and $a_v$ as fitting parameters, we can approximate actual form factor with any precision.

In the application to double photon exchange contribution, the modification \eqref{eq:FFsQED} corresponds to simply multiplying the  \emph{integrand} of sQED amplitudes by $\F(q_1^2)\F(q_2^2)$. Such a multiplication respects gauge invariance and clearly implements the properties that we discussed in the end of the previous section.

\begin{figure}
    \centering
    \includegraphics[width=0.8\linewidth]{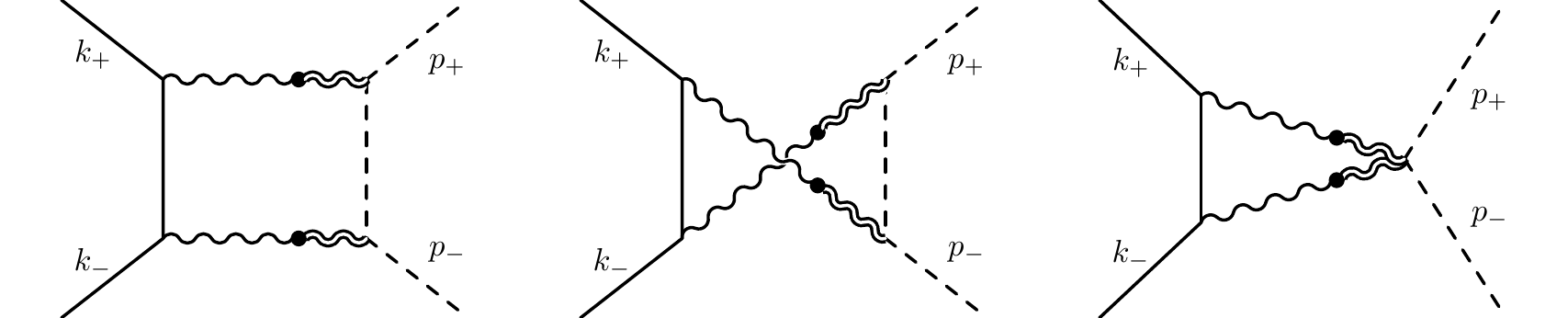}
    \caption{Diagrams for double photon exchange contribution within our model.}
    \label{fig:FFsQED}
\end{figure}

Thus, we are to calculate the sum of diagrams depicted in Fig. \ref{fig:FFsQED}.

%$$
%\vcenter{\hbox{\includegraphics{Figs/Feynmf/GVDff}}}=
%\vcenter{\hbox{\includegraphics{Figs/Feynmf/GVDff1}}}=\F(q^2)\times
%\vcenter{\hbox{\includegraphics{Figs/Feynmf/GVDff2}}}\,,
%$$
\section{Technical details and results}
First, simple considerations show that there are no collinear divergences in diagrams in Fig. \ref{fig:FFsQED} in the limit of zero electron mass, $m_e=0$. This was already underlined in Ref. \cite{Arbuzov:2020foj} for the case of sQED, and our modification preserves this property. Thus we put electron mass to zero\footnote{This is valid up to power corrections in $m_e$, which are negligible for our case.}. Then the contribution of seagull diagram (the rightmost diagram in Fig. \ref{fig:FFsQED}) can be shown to vanish. Thus, we are left with the box and cross-box diagrams which differ by the replacement $t\leftrightarrow u$.

It is convenient to split the contribution into two pieces: that in the standard sQED approach and the additional contribution
\begin{equation}
    \delta^V=\delta^V_{\text{sQED}}+\delta^V_{\text{FF}}
\end{equation}
We use standard technique of multiloop calculations, based on the reduction to master integrals \cite{Tkachov1981,ChetTka1981}. 

Let us start with sQED contribution which is infrared divergent. To regularize this divergence, we use dimensional regularization, $d=4-2\epsilon$. Therefore, we can not directly compare our result of $\delta^V_{\text{sQED}}$ with that of \cite{Arbuzov:2020foj} as that paper used regularization with fictitious photon mass. In order to make a comparison, we had to calculate the soft real radiation contribution coming from ISR-FSR interference also in dimensional regularization. Fortunately, the corresponding formulae have been derived in Ref. \cite{Lee:2020zpo}. Adding up these two contributions, we obtain
\begin{multline}
    \delta_{\text{sQED}}^V+\delta^S=\frac{\alpha}{\pi}\Bigg[
%dvs
    2 \ln \left(\frac{1+\beta  c}{1 - \beta  c}\right) \ln {\frac{\sqrt{s}}{2 \omega_0}}+\frac{1}{2}  \ln \left(\frac{1+\beta  c}{1 - \beta  c}\right)\ln \left(\frac{1+\beta}{1-\beta}\right)\\
    +\frac{(1 + \beta  c)^2 }{\beta ^2\, \left(1-c^2\right)}\ln (1 + \beta  c) \ln \left(\frac{1 + \beta  c}{1-\beta ^2}\right)+2 \mathrm{Li}_2\left(\frac{(1+c) \beta }{1+\beta}\right)+2 \mathrm{Li}_2\left(\frac{(1+c) \beta }{1+c \beta}\right)\\
    +\frac{c \ln ^2\left(1-\beta ^2\right)}{2 \beta\,(1 - c^2)}
    +\frac{\left(1+2 \beta  c+\beta ^2\right) }{\beta ^2\, \left(1-c^2\right)}\left(\mathrm{Li}_2\left(\frac{1+2 \beta  c+\beta ^2}{\beta ^2-1}\right)+\frac{\pi ^2}{12}\right)\\
    +\frac{\left(1+\beta ^2\right) c}{\beta ^2\, \left(1-c^2\right)} \left( \mathrm{Li}_2\left(\frac{\beta -1}{\beta +1}\right)+\frac14\ln ^2\left(\frac{1+\beta}{1-\beta }\right)+\frac{\pi ^2}{12}\right)
%dvs/
    \Bigg]
     -\left(c\to-c\right)\label{eq:deltaSQED}
\end{multline} 
which is to be compared with the sum of Eqs. (5) and (7) of Ref. \cite{Arbuzov:2020foj}. Note that our result is somewhat simpler. Nevertheless, we find perfect numerical agreement with arbitrarily high precision, which unambiguously signals the possibility to reduce the two results to each other using identities between dilogarithms\footnote{In principle, this is a routine task if one uses symbol map \cite{Duhr:2011zq}, but we omit this check as numerical agreement is pretty convincing.}.

Let us now pass to the additional contribution. This contribution is obtained by multiplying the sQED integrand by the factor \linebreak $\left[\F(q_1^2)\F(q_2^2)-\F(q^2)\right]/\F(q^2)$. This factor vanishes when either of $q_{1,2}$ goes to zero. Therefore, this additional contribution is IR finite. By power counting, the additional contribution is, of course, also UV finite. Therefore, for its calculation, it convenient to use master integrals which remain finite at $d=4$. These master integrals can be calculated directly using Feynman parametrization in terms of dilogarithm function. However, it is not trivial to obtain compact explicit expressions valid for all possible parameters and we prefer to leave a one-fold representation for two most complicated functions ($g_4$ and $g_8$, see below). Then we obtain

\begin{multline}
\delta^V_{\text{FF}}=\mathrm{Re}\frac{2\alpha}{\pi s \beta^2(1 - c^2)\F(s)} \sum_{i,k}a_ia_k\Bigg\{
%dff
    \frac{t^2\,(g_4-2g_2)}{s}- \left(2T^2+s\,t +2\Lambda_i t\right)\frac{g_8}{s}\\
    +\frac{t^2 g_1}{s-\Lambda_i} 
    -t\,g_5-\frac{2 \left(2T^2+s\,t+\Lambda_i t\right)\left(g_7-g_3\right)}{ \left(s-\Lambda_i\right)} - \frac{\left(2T^2+s\,t\right)g_6 }{ s-\Lambda_i}
%dff/
    \Bigg\}-(t\leftrightarrow u)\,,
    \label{eq:deltaFF}
\end{multline}
where $\F(s)$ is defined in Eq. \eqref{eq:FF}, $t=m_\pi^2-(1-\beta c)s/2,\quad u=m_\pi^2-(1+\beta c)s/2,\ T=m_\pi^2-t, c=\cos \theta$, and  the functions $g_1,\ldots , g_8$ are defined in the Appendix. We attach the corresponding \texttt{Mathematica} and \texttt{C++} files to the arXiv submission.

\section{Form factor fits and numerical results for $\delta^{V}_{\text{FF}}$}

\begin{figure}
    \centering
    \includegraphics[width=.6\linewidth]{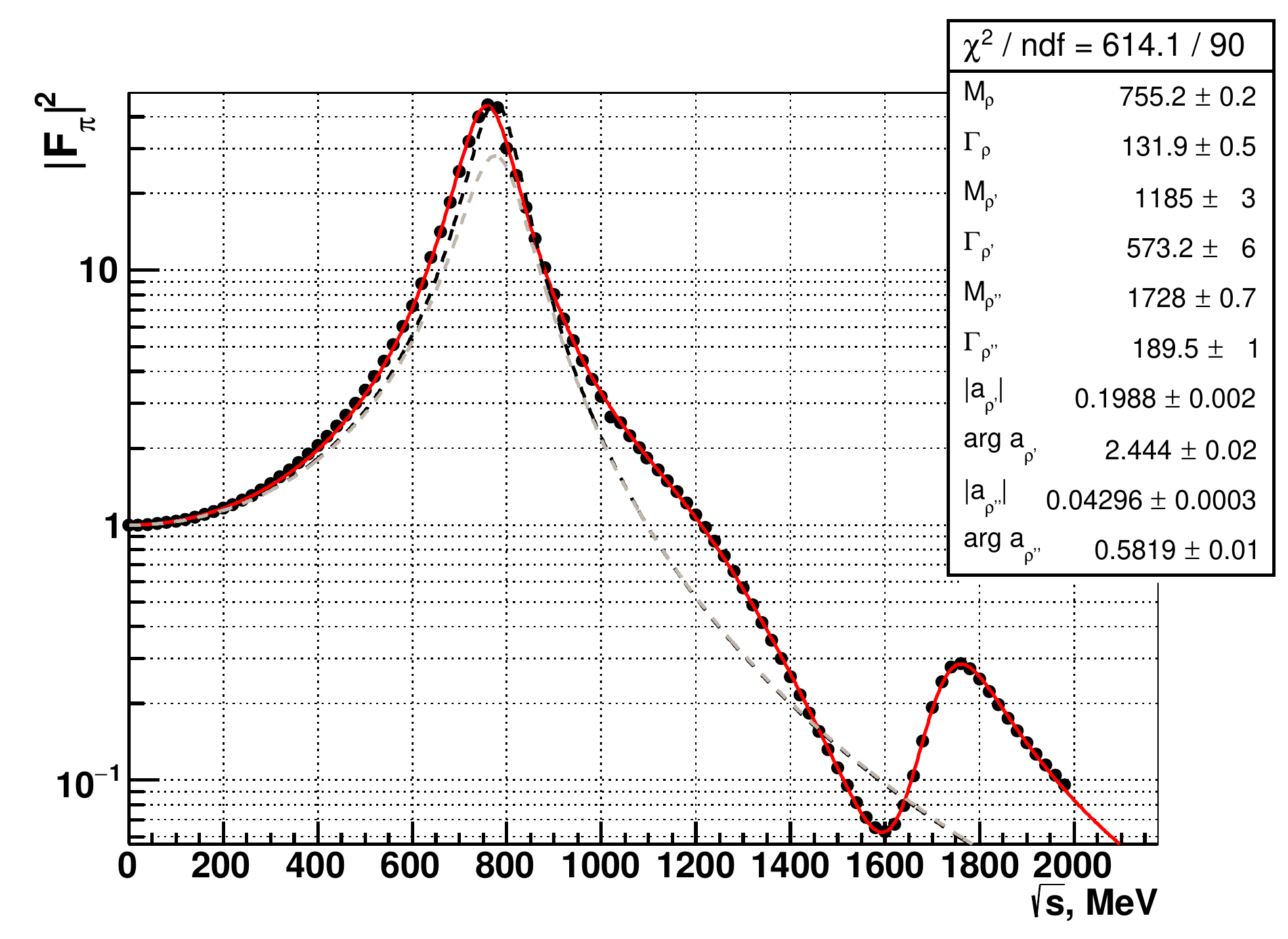}
    \label{fpifit} 
    \caption{Pion form factor function used for the $\delta^{V}_{\text{FF}}$ calculation.
        Points correspond to the values of the model with sum of the
        Gounaris-Sakurai resonance parametrization, which is used to describe 
        experimental $e^+e^-\to\pi^+\pi^-$ cross-section. Dashed lines
        correspond to single BW$_\rho$ (see description in the text). 
    }
    %\vspace{-0.3cm}
\end{figure}

\begin{figure}
    \centering
    \includegraphics[width=.495\linewidth]{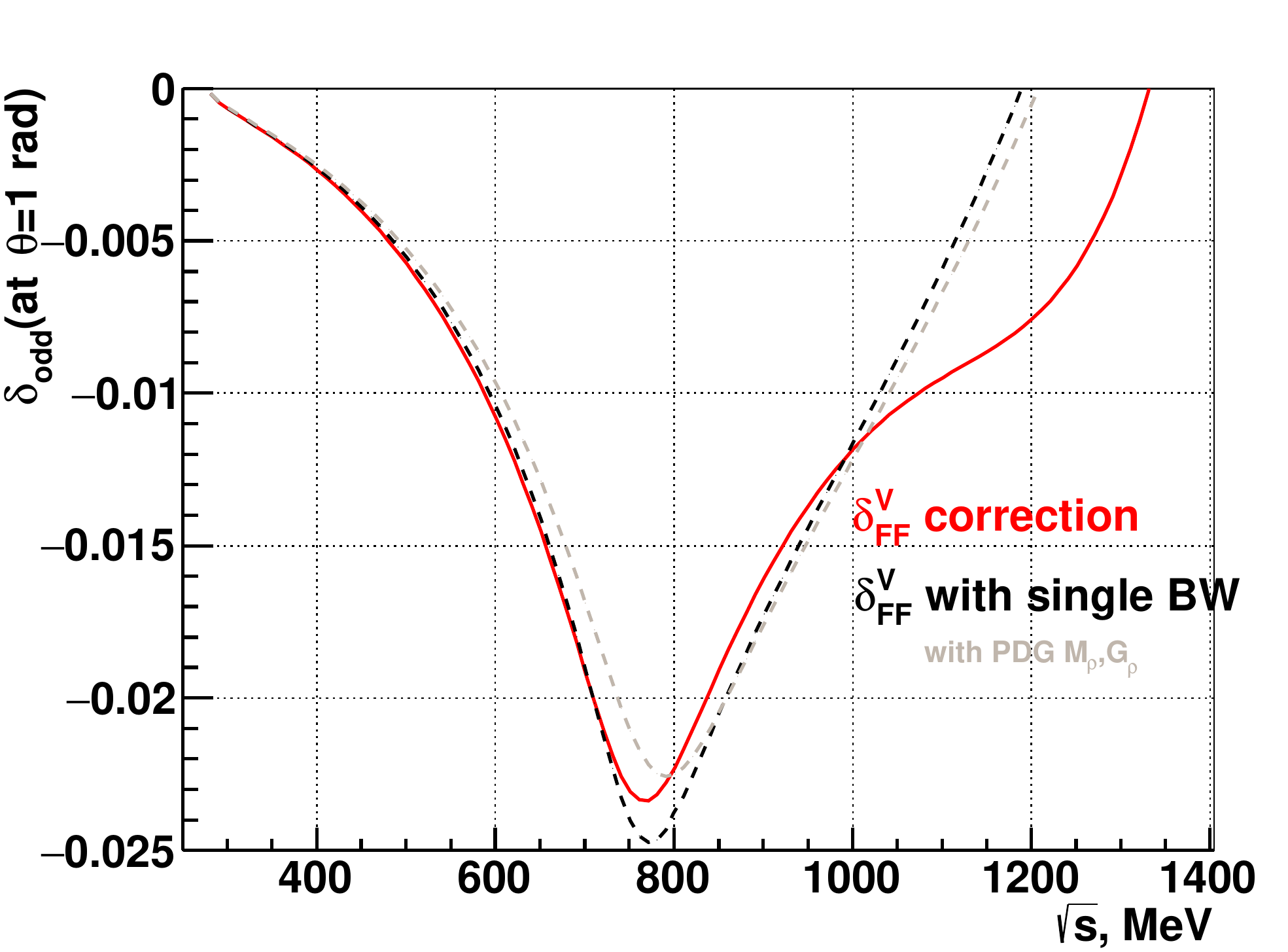}
    \includegraphics[width=.495\linewidth]{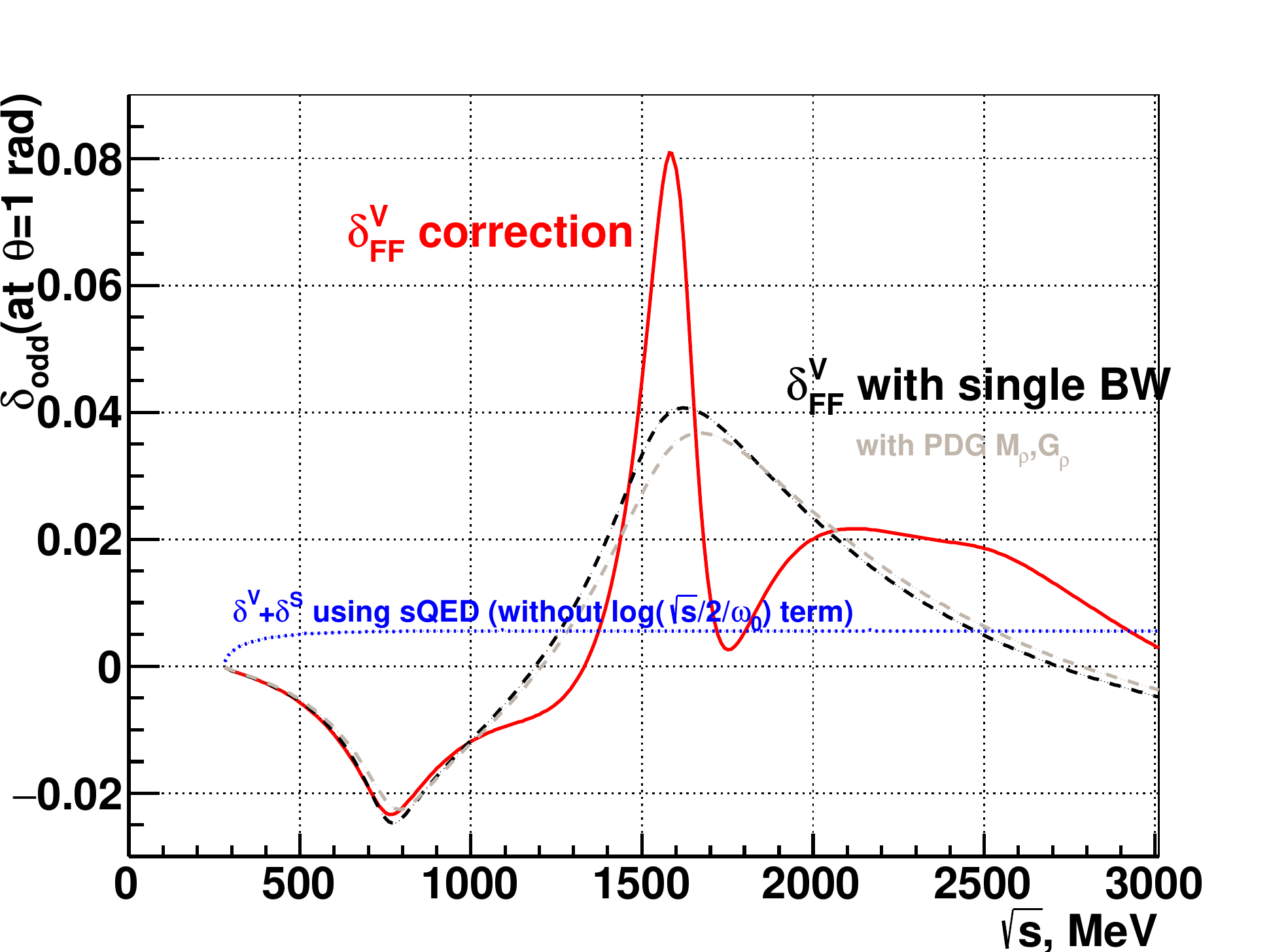}
    \caption{Dependence of $\delta^{V}_{\text{FF}}$  at $\theta=1$~rad with energy.
        Red line function corresponds to the use of form factor with sum of three $\rho,\rho',\rho''$ resonances.
        Black dashed line with using single BW$_\rho$ form factor such that $|F(m_\rho^2)|^2$ match experimental value, 
        gray dashed line by using PDG's $m_\rho,\Gamma_\rho$ values.
        Blue dotted line correspond to the sum $\delta^V_{\text{sQED}}+\delta^S$ within sQED approach (without $\ln\frac{\sqrt{s}}{2\omega_0}$ term in Eq. \eqref{eq:deltaSQED}). }
    \label{fig:deltaff} 
    %\vspace{-0.3cm}
\end{figure}

\begin{figure}
%    \begin{minipage}[t]{.49\textwidth}
%        \centering
%        \includegraphics[width=.99\linewidth]{deltavffvsG-eps-converted-to.pdf}
%        \caption{$\delta^{V}_{\text{FF}}(m_\rho)$ with single $BW_\rho$ formfactor parametrization
%            at $\theta=1$~rad versus different $\Gamma_\rho$ values. }
%        \label{deltaffvsG} 
%        %\vspace{-0.3cm}
%    \end{minipage}
%    \quad
%    \begin{minipage}[t]{.49\textwidth}
        \centering
        \includegraphics[width=.7\linewidth]{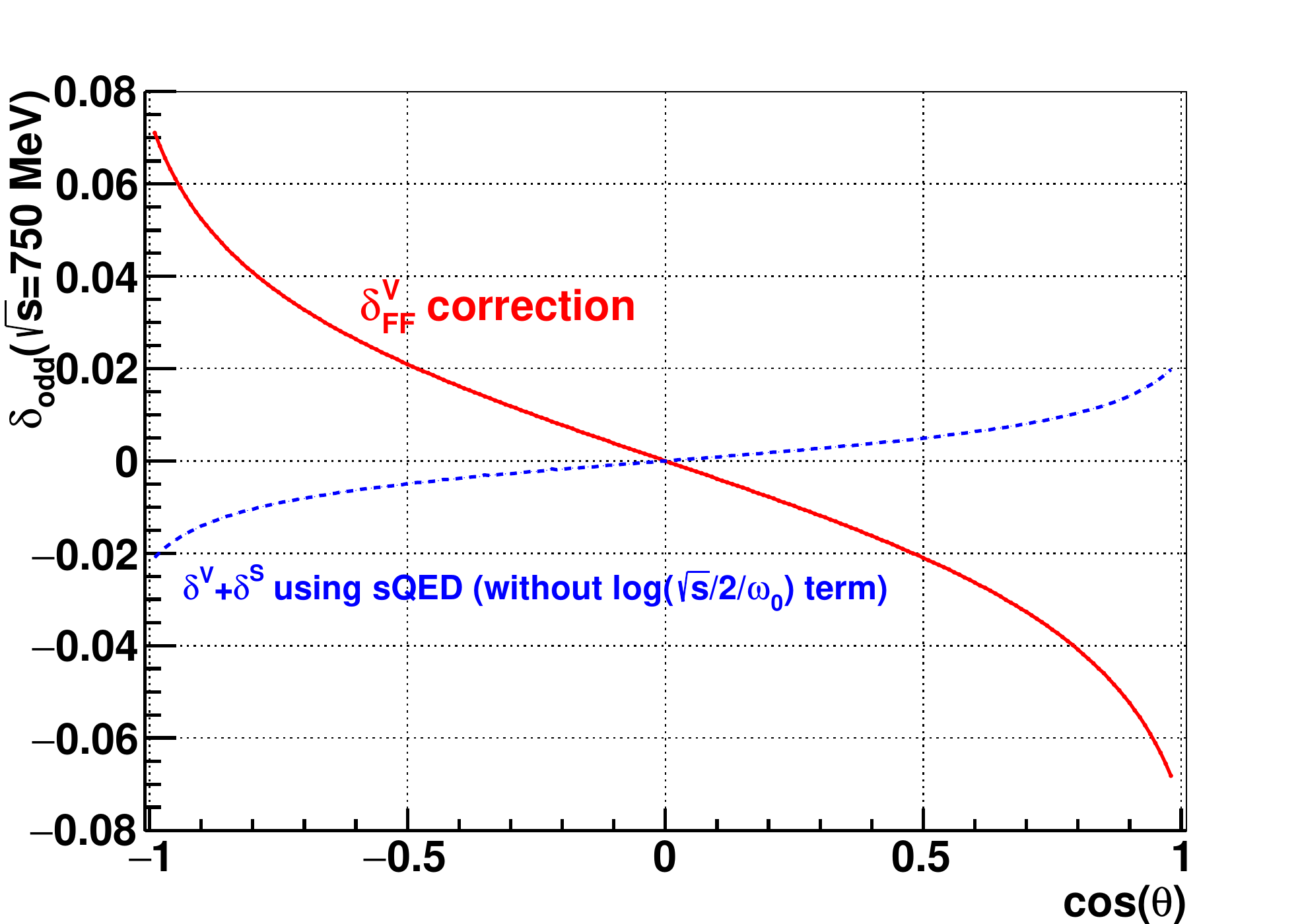}
        \caption{$\delta^{V}_{\text{FF}}$ correction versus $\cos \theta$ at
            $\sqrt{s}=750$~MeV.
            Blue dotted line correspond to virtual and soft correction $\delta^V_{\text{sQED}}+\delta^S$
            from Eq. \eqref{eq:deltaSQED} (without $\ln\frac{\sqrt{s}}{2\omega_0}$ term) within sQED approach.}
        \label{fig:deltaffvsc} 
%    \end{minipage}%
\end{figure}
For the numerical evaluation of the obtained correction, we use the pion form factor measured in CMD-2 experiments, see Refs. \cite{CMD-2:2006gxt,CMD-2:2005mvb}. As explained in those papers, the measured form factor is accurately described by the Gounaris-Sakurai parametrization, which we refit by the sum of three Breit-Wigner (BW) functions, Eq.~\eqref{eq:FF} with $n=3$. The result of this fit is shown in Fig.~\ref{fpifit}. As one can see from this figure the parametrization~\eqref{eq:FF} with three BW functions describes the form factor rather well with typical deviation being less than 5 percent. In the same figure we also show two simpler approximations with a single Breit-Wigner function: one, corresponding to both $m_\rho$ and $\Gamma_\rho$ taken from PDG (gray curve), another, with the width reduced by a factor $\approx 1.26$  (black curve), so that the peak value of the BW function coincides with that of the actual form factor.

Using these Breit-Wigner parametrizations, we have calculated the
correction $\delta^{V}_{\text{FF}}$, Eq. \eqref{eq:deltaFF}, as a function of
energy and angle. The energy dependence of $\delta^{V}_{\text{FF}}(s)$ at
$\theta$=1 rad is shown in Fig. \ref{fig:deltaff}. From the left plot of Fig. \ref{fig:deltaff} one can see that the value of $\delta^{V}_{\text{FF}}$ only slightly depends on the parametrization of the form factor up to $\sqrt{s}\approx 1$ GeV. 
On the right plot the energy dependence is shown for a wide energy range. Note that in the region around the rho meson mass, the contribution  $\delta^{V}_{\text{FF}}$ is a few times larger than the additive constant at the large logarithm in $\delta_{\text{sQED}}^{V}+\delta^S$ in Eq. \eqref{eq:deltaSQED}. One can observe a strong dependence on the form factor parametrization when $\sqrt{s}>1$ GeV. This might be considered as the indication that in this region our model loses its validity.

A typical angular dependence of $\delta^{V}_{\text{FF}}$ is shown in Fig. \ref{fig:deltaffvsc}. Note that this dependence is quite close to linear when $|\cos \theta|<0.5$.

\section{Comparison with the experiment}

The charge asymmetry in the $\pi^+\pi^-$ channel was studied with the
CMD-3 detector~\cite{Khazin:2008zz,pipipaper}.
This analysis includes stable particle separation, precise fiducial volume determination, theoretical precision of radiative corrections, etc. 
Good understanding of the observed angular distributions is important for investigation and controlling the systematic effects in the fiducial volume determination. The magnitude of the obtained correction $\delta^{V}_{\text{FF}}$ to the differential cross section is about $2.5\cdot 10^{-2}$ at $\theta$=1 rad and $m_\rho$ peak which should be compared to a few ppt level of a systematic precision requirement for the $e^+e^-\to\pi^+\pi^-$ cross section measurements as it was described in the introduction.
The measured asymmetry is defined as the difference between the  detected numbers of events to the forward and backward regions of the detector, 
\begin{equation}
    A=\frac{N_{\theta<\pi/2}-N_{\theta>\pi/2}}{N_{\theta<\pi/2}+N_{\theta>\pi/2}}\,.
\end{equation}
The following event selection criteria on pion momenta, acceptance angles and collinearity angle were used in the experiment:
\begin{gather}
    p^{\pm}>0.45E_{\text{beam}},\quad  1<\theta=\tfrac12( \pi -\theta^++\theta^-)<\pi - 1\text{ rad},\nonumber\\ 
    \left||\phi^+-\phi^-|-\pi\right|<0.15\text{ rad},\quad  |\theta^+ + \theta^- - \pi|<0.25 \text{ rad},\label{eq:selcrit}
\end{gather}
where $E_{\text{beam}}=\sqrt{s}/2$.
 
These numbers were corrected for detector effects like inefficiencies, resolution smearing,
etc, so that the measured asymmetry can be directly compared with the predictions obtained with Monte-Carlo generator. The calculated $\delta^{V}_{\text{FF}}$ correction was introduced into the MCGPJ
generator~\cite{Arbuzov:2005pt}, which already includes exact
NLO to $e^+e^- \rightarrow \pi^+\pi^-$ with point-like pion and
additional ISR jets along beam axis using the structure functions approach to
partially account the higher order logarithmically enhanced effects.

\begin{figure}[!b]
    \centering
    \includegraphics[width=.495\linewidth]{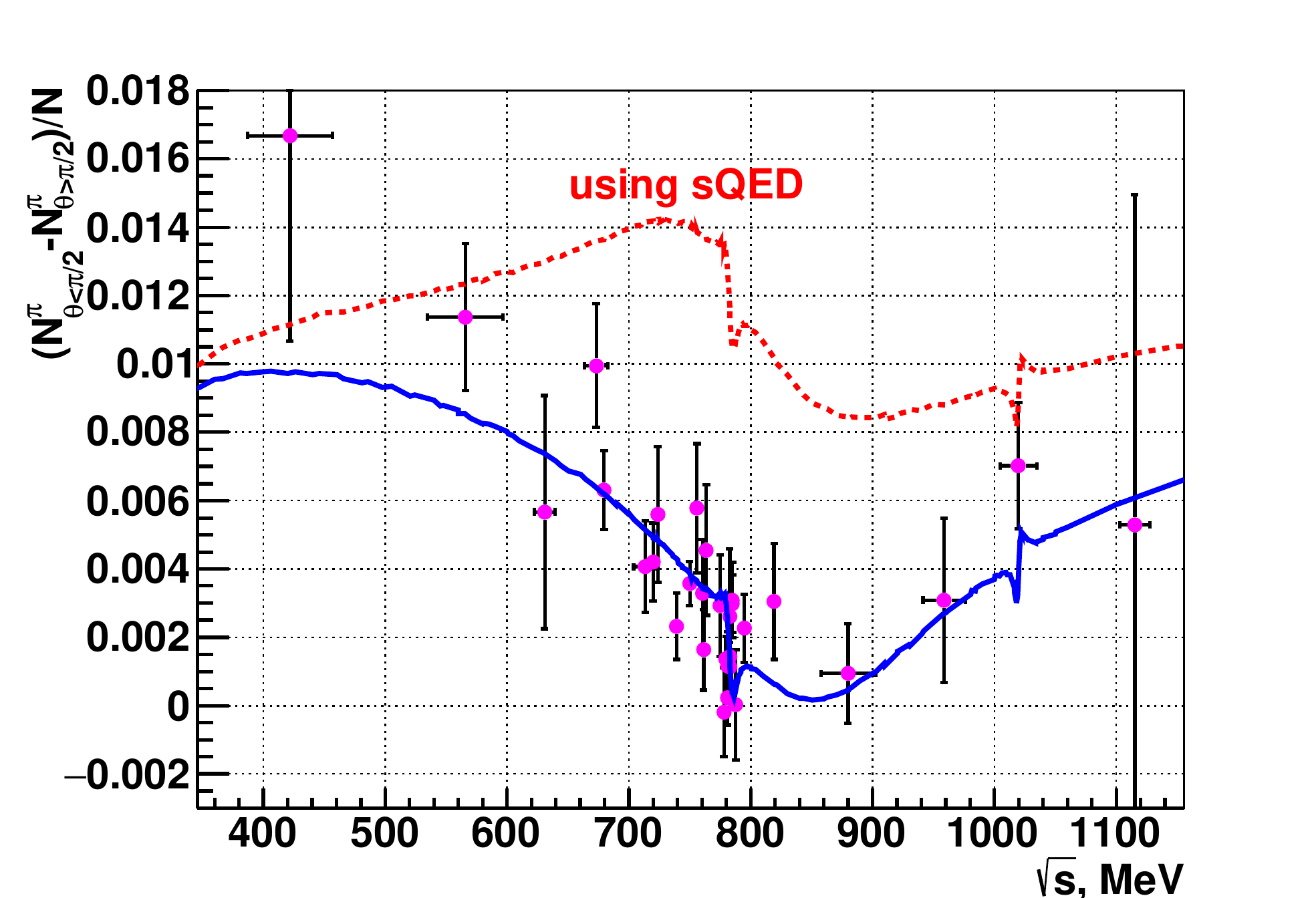}
    \includegraphics[width=.495\linewidth]{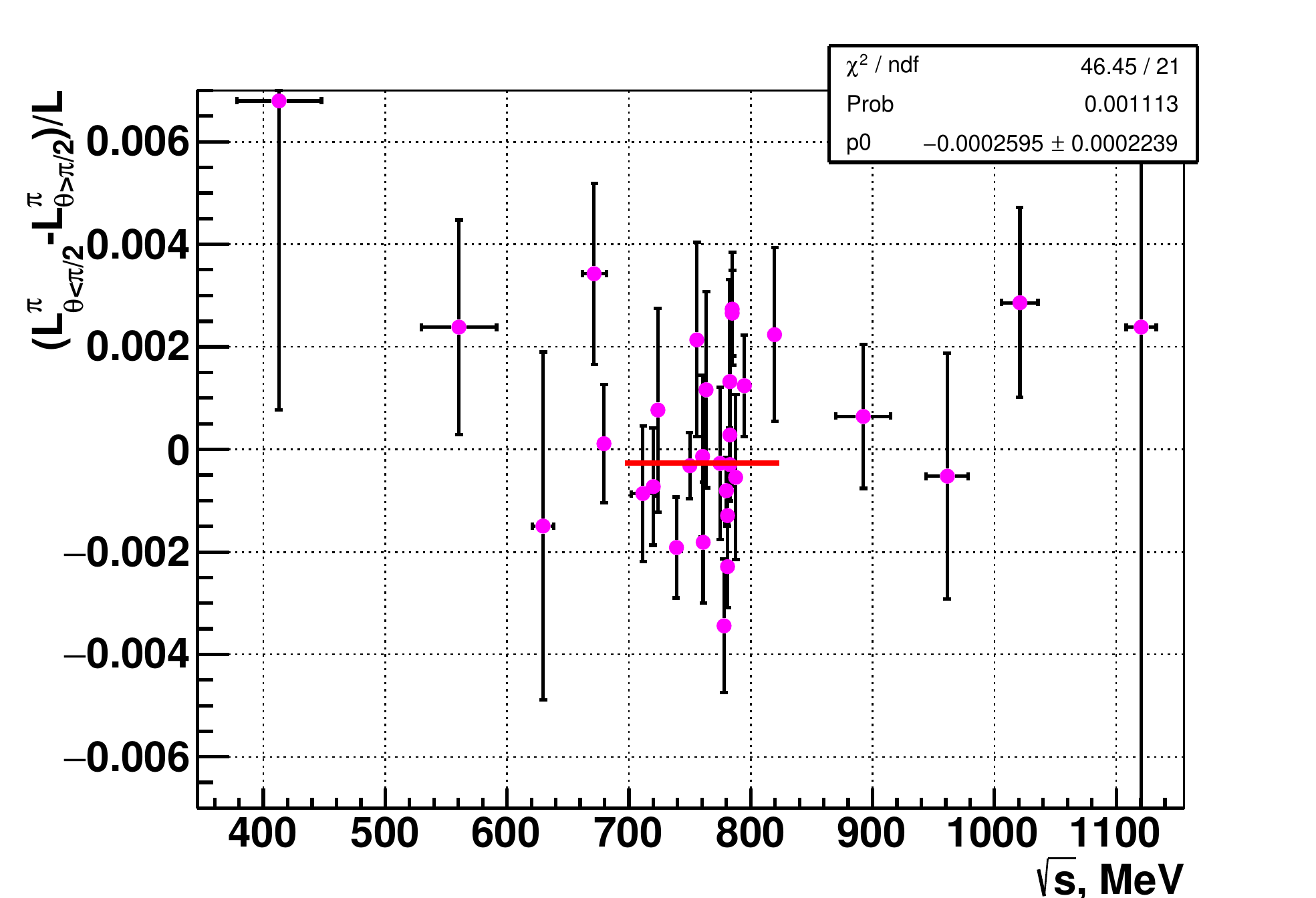}
    \caption{Left plot: the measured asymmetry in $\pi^+\pi^-$ at the CMD-3 in comparison with
        prediction based on commonly used sQED approach (red dotted line),
        and our present approach (blue line). Right plot:
        difference between measured number of events and prediction based on
        the present calculation.}
    \label{asym} 
    %\vspace{-0.3cm}
\end{figure}

The selection criteria \eqref{eq:selcrit} can be roughly translated to photon energy cutoff $\omega_0 \lesssim 0.2 E_{\text{beam}}$, which corresponds to the magnitude $\sim0.02$ for the first term in Eq.~\eqref{eq:deltaSQED} at $\theta=1$ rad. This leads to $\sim 0.01$ contribution to the asymmetry from the ISR-FSR interference within these experimental cuts.

The obtained experimental result together with theoretical prediction are shown in Fig.~\ref{asym}. The points correspond to the overall statistics collected to the present moment with the CMD-3 detector at $\sqrt{s}<1$ GeV.
The dotted red line corresponds to the prediction of the conventional approach based on sQED calculations. The discrepancy of this prediction with experimental data is much larger than statistical precision of the latter (more than 40$\sigma$).
%has the magnitude about $0.01$ in the $\rho$-peak region with $\pm2.2\cdot 10^{-4}$ statistical precision
The theoretical result for the asymmetry obtained in the present paper is shown by the blue line in Fig.~\ref{asym}. As it is seen, it agrees well with the measured asymmetry. The relative difference between the measured asymmetry and the theoretical result is well within statistical precision, as shown on the right plot of this figure. The average value of this difference over the range $\sqrt{s}=700\div 820$ MeV is  $(-2.6\pm 2.2)\cdot 10^{-4}$ which is to be compared to  $(104\pm 2.2)\cdot 10^{-4}$ difference for conventional approach.

\section{Conclusion}

In the present paper we consider the charge asymmetry in the process of $\pi^+\pi^-$ pair production in the electron-positron annihilation. 
We demonstrate that the conventional approach based on sQED calculations is not valid  in the region of $\rho$ peak because of a strong energy dependence of the pion form factor in this region. 
As a consequence, this approach is in clear contradiction with the experimental data for the asymmetry obtained with CMD-3 detector. 
We describe a simple and natural approach to the calculation of the asymmetry based on generalized vector dominance and explain why it appropriately describes the energy region around $\rho$ mass. 
The comparison of the results of this approach with experimental data shows a remarkable agreement.

The obtained result shows the importance of the appropriate choice of
the model for the calculation of the radiative corrections for the
$\pi^+\pi^-$ channel.  It shows the necessity to critically revisit
other calculations withing sQED approach. We note that this approach
was used in Ref.   \cite{Campanario:2019mjh} when considering the NNLO
corrections for $\pi^+\pi^-$ total cross section measurements within the ISR method.

\paragraph{Acknowledgement} We are grateful to A. Bondar and A. Milstein for the interest to the work and valuable discussions.

\appendix

\section{Functions $g_1,\ldots, g_8$} 

\begin{align}
%    %J12
%    g_1&=\intop_0^1 dx 
%%f1i
%    \frac{\ln \left(-\frac{m_\pi^2 x^2}{s\, \bar{x}}+i0\right)}{s\, \bar{x}+m_\pi^2 x^2}
%%f1i/
%    =
%%f1a
%    g_1/s\\
%%f1a/
    g_1&=
%F1a
    \frac{1}{\beta}\left[2\text{Li}_2\left(-\tfrac{1+\beta}{1-\beta}\right)+ \tfrac{1}{2} \ln^2 \left(\tfrac{1+\beta}{1-\beta}\right)+i\pi \ln \left(\tfrac{1+\beta}{1-\beta}\right)+\tfrac{\pi ^2}{6}\right]
%F1a/
    ,\\
%    %J13
%    f_2&=\intop_0^1 dx
%%f2i
%    \frac{ \ln {\frac{\Lambda_i\bar{x}+m_\pi^2 x^2}{\bar{x}\,\left(\Lambda_i-s\right)}}}{s\,\bar{x}+m_\pi^2 x^2}\,
%%f2i/
%    =
%%f2a
%    g_2/s
%%f2a/
%    \nonumber\\
    g_2&=
%F2a
\frac{1}{\beta}\bigg[\ln {\tfrac{1+\beta _i}{1-\beta _i}}\ \ln {\tfrac{(1-\beta ) \left(\beta _i+\beta\right)}{(1+\beta) \left(\beta _i-\beta \right)}}
+\text{Li}_2\left(\tfrac{(1+\beta) \left(1-\beta _i\right)}{(1-\beta) \left(1+\beta _i\right)}\right)
-\text{Li}_2\left(\tfrac{(1-\beta) \left(1-\beta _i\right)}{(1+\beta) \left(1+\beta _i\right)}\right)\Big]
%F2a/
    \,,\\
%    %J15
%    f_{3}&=\intop_0^1{dx}
%%f3i
%    \frac{\ln {\frac{\Lambda_i \bar{x}+T\,x}{\Lambda_i \bar{x}+m_\pi^2 x^2}}}{x\, \left(m_\pi^2 \bar{x}-t\right)}
%%f3i/
%    =
%%f3a
%    g_3/T
%%f3a/
%    \nonumber\\
    g_3&=
%F3a
    \text{Li}_2\left(\tfrac{t}{m_\pi^2}\right)-\text{Li}_2(1-p_i)-\text{Li}_2(1-q_i)+\tfrac{\pi ^2}{6}
%F3a/
    ,\\
%    %J19
%    f_4&=
%%f4a
%    g_4/s
%%f4a/
%    \nonumber\\
    g_4&=\intop_0^{1}dx 
%F4i
    \frac{\ln \left[\frac{B-s \bar{x}+\Lambda_i-\Lambda_k}{B+s \bar{x}+\Lambda_i-\Lambda_k}\right]+\ln \left[\frac{B-s \bar{x}+\Lambda_k-\Lambda_i}{B+s \bar{x}+\Lambda_k-\Lambda_i}\right]}{B/s}
%F4i/
   ,\nonumber\\
    B&=
%F4B
    \sqrt{s^2 \bar{x}^2-2 \left(\Lambda_i+\Lambda_k\right) s \bar{x}+\left(\Lambda_k-\Lambda_i\right)^2-4 m_\pi^2 s x^2}
%F4B/
    ,\quad \bar{x}=1-x,\\
%    %J20
%    f_5&=\operatorname{Re}\intop_0^{1}dx
%%f5i
%    \frac{ \ln {\frac{\Lambda_i\bar{x}+\Lambda_k x}{\Lambda_i\bar{x}+\Lambda_k x-s x\bar{x}}}}{s x\bar{x}}
%%f5i/
%	=
%%f5a
%    g_5/s
%%f5a/
%    \nonumber\\
    g_5&=
%F5a
    \ln \left(\tfrac{R+\Lambda _i-\Lambda _k+s}{R+\Lambda _i-\Lambda _k-s}\right) \ln \left(\tfrac{R-\Lambda _i+\Lambda _k-s}{R-\Lambda _i+\Lambda _k+s}\right)
%F5a/
    , \quad
    R=
%F5R
    \sqrt{\left(s-\Lambda _i-\Lambda _k\right)^2-4 \Lambda _i \Lambda _k}
%F5R/
    \,,\\
%    %J_21
%    f_6& = 
%%f6a
%    -\frac{s\,T}{16} g_6
%    -\frac{s}{16 t} \left[s\,T+4t\, U\right] \ln \left(\tfrac{T}{m_{\pi}^2}\right)+\frac{s\,(t-u)}{16} \left[\ln \left(\tfrac{s}{m_\pi^2}\right)-i\pi\right]
%%f6a/
%    \,,\nonumber\\
    g_6&=
%F6a
    2 \text{Li}_2\left(\tfrac{t}{m_\pi^2}\right)+\frac{1}{2} \ln ^2\left(\tfrac{T^2}{m_\pi^2s}\right)+i\pi \ln\left(\tfrac{T^2}{m_\pi^2s}\right)+\tfrac{\pi ^2}{6}
%F6a/
    \\
%    %J23
%    f_{7}&%(s,t,m_\pi,\Lambda_1,\Lambda_2%)
%    = \intop_0^1 dx 
%%f7i
%    \frac{s\, \left(t-u \bar{x}+m_\pi^2 x\right) }{4 x\, \left(m_\pi^2 \bar{x}-t\right)}\bigg[
%    \frac{\ln {\frac{\Lambda_i \bar{x}+\left(m_\pi^2-t\right)x}{\bar{x}\, \left(\Lambda_i-s\right)}}}{s \bar{x}+ \left(m_\pi^2-t\right)x}-\frac{\ln {\frac{\Lambda_i \bar{x}+m_\pi^2 x^2}{\bar{x}\, \left(\Lambda_i-s\right)}}}{s \bar{x}+m_\pi^2 x^2}\bigg]
%%f7i/
%    =
%%f7a
%    \frac{s}{4}f_3-\frac14g_7
%%f7a/
%    \nonumber\\
    g_7&= 
%F7a
    \ln \left(1-\tfrac{s}{\Lambda _i}\right) \ln \left(\tfrac{T^2}{m_\pi^2 s}\right)+ \text{Li}_2\left(1-\tfrac{\Lambda _i}{s}\right)+\frac{1}{2} \ln ^2\left(\tfrac{\Lambda _i}{s}\right)+\tfrac{\pi ^2}{6}
%F7a/
    ,\\
    %J28
%    f_{8}&=
%%f8a
%    g_8/(s\,T)
%%f8a/
%    \nonumber\\
    g_8&=\intop_0^{\infty}dy\,
%F8i
    s\,T\frac{\ln \left[\frac{A+\Lambda_i-\Lambda_k-s}{A+\Lambda_i-\Lambda_k+s}\right]+\ln \left[\frac{A+\Lambda_k-\Lambda_i-s}{A+\Lambda_k-\Lambda_i+s}\right]}{A\, \left(m_\pi^2+ T\, y\right)}
%F8i/
    ,\nonumber\\
    &A=
%F8A
    \sqrt{\left(\Lambda_i-\Lambda_k\right)^2-2 s\, \left(\Lambda_i+\Lambda_k+2\Lambda_i\Lambda_k y\,(1+y)/m_\pi^2\right)+s^2}
%F8A/
    \\
    T&=m_\pi^2-t\,,\quad  \beta_i=\sqrt{1-4m_\pi^2/\Lambda_i}\,,
    \quad p_i=\frac{T\left(1+\beta_i\right)}{2 m_\pi^2}\,,
    \quad q_i=\frac{T\left(1-\beta_i\right)}{2 m_\pi^2}\,.
\end{align}
\bibliographystyle{elsarticle-num} 
%\bibliography{pipiAsym}

\end{document}